\documentstyle[12pt]{article}
\begin{document}
\begin{center}
{\bf Lift of degeneracy of Landau levels of 2D electron gas by point-like impurities}

\bigskip
 A. M. Dyugaev$^{1,2}$, P. D. Grigoriev$^{1,3}$,
 Yu. N. Ovchinnikov$^{1,2}$

\bigskip
\small{
{\it
$^1$ L.D. Landau Institute for Theoretical Physics \\
Chernogolovka, Moscow region, 142432, Russia

$^2$ Max-Planck-Institut for the Physics of \\
Complex Systems, Dresden D-01187 Germany

$^3$ High Magnetic Field Laboratory, MPI-FRF and CNRS, BP166,
Grenoble, F-38042, France, e-mail: pashag@itp.ac.ru }}
\end{center}

\vspace{1cm}

\small{We study the density of states of two-dimensional electron
gas in magnetic field with scattering on point-like impurities that are uniformly distributed
along the sample. We show that the electron-impurity interaction completely
lifts the degeneracy of the Landau levels even at low impurity
concentration. This statement is in contradiction with the previous
results obtained in the unphysical approximation of two-dimensional
impurities. There is a large region of the electron energy $\omega $, counted from
the Landau level, where the density of states $\rho (\omega)$ is inversely proportional
to $|\omega|$ and proportional to the impurity concentration.
Our results are applicable to different 2D electron systems as
heterostructures, inversion layers and interfaces of condensed media.

\bigskip

PACS: 72.10.Fk}

\vspace{0.5cm}

{\bf 1.} Two-dimensional (2D) electron systems are realized on interfaces
of two condensed media. The typical examples of such systems are heterostructures \cite{QHE},
electrons on the helium surface \cite{Sh} and in inversion layers.
In magnetic field the spectrum of the 2D electrons is discrete and
infinitely degenerate. We consider the question of removal of this degeneracy
due to interaction of electrons with point-like impurities.
There is an opinion that at low surface concentration $n_s$ of these impurities
the infinite degeneracy the Landau levels is lifted only partially \cite{Baskin}.
More precisely if $n_s<\frac{1}{2\pi l^2_H}$ the degeneracy of the Landau levels is
$\frac{S}{2\pi l^2_H}-N$. Here $N=n_s S$ is the total number of impurities, $S$
is the surface accessible for electrons, $l_H=\hbar c/eH$ is the magnetic length.
In this case $N$ electrons separate from each Landau level and form the impurity band.
The proof of this statement \cite{Baskin} is based on the possibility to determine the wave function of electrons
on the Landau level in such a way that it is zero at the points where the impurities are
situated. If the impurities are point-like they do not influence the spectrum of
 $\frac{S}{2\pi l^2_H}-N$ electron states. However, such a conclusion need the
 surface impurity density $n_S$ to be determined.
The problem is that the 2D electron systems are usually \underline{open}.
The electron motion along the $z$ axis which is perpendicular to the conducting planes is
characterized by the wave function $\varphi (z)$ which is confined on the macroscopic scale $z_{\infty}$
that depends on the geometry of sample. The typical wave function
$\varphi (z)$ has the form \cite{Sh,QHE}
\begin{equation}
\label{1}
\varphi^2(z)=\frac{1}{2z_0}z^2_{*}e^{-z_{*}} \quad
\mbox {\rm at} \quad  0<z<z_{\infty}
\end{equation}
\begin{displaymath}
\varphi (z)=0 \quad \mbox {\rm at} \quad z<0, \quad
z_{*}\equiv z/z_0
\end{displaymath}
Electrons can move in the region $0<z<z_{\infty}$ (not only in $z\le
z_0$).
Hence, even at low bulk concentration $n_{imp}$ of impurities
their surface density $n_S=n_{imp}z_{\infty}$ is very large
(the effective interaction of electrons with an impurity now depends on the distance
of this impurity from the conducting plane). Usually one can put $n_S=\infty$.
Hence, the infinite degeneracy of the Landau levels is lifted completely
by the point-like impurities.
 In other words the number of impurities is small in the region
$z\le z_0$, but large in the region $z< z_{\infty}: C_{\infty}=C_0
\frac{z_{\infty}}{z_0} \gg 1$. This situation usually takes place
on experiment \cite{Sh}, \cite{QHE}. For electrons on the liquid helium surface
the heavy atoms of helium vapor play the role of impurities.
Below we calculate the electron density
 $\rho$ at low impurity concentration $C_0\equiv n_{imp} 2\pi l^2_Hz_0\ll 1$.
\medskip

{\bf 2.} The density of states $\rho (E)$ is related to the imaginary part of the electron Green's
function $G(r,r',E)$ by the relation $\rho (E)
=\frac{-1}{\pi} Im G (E)$, where $G(E)=\int d^3 r G(r,r,E)$ \cite{AGD}.
Since the electron system is uniform along the conducting $(x,y)$ plane, the Green's
function does not depend on $(x,y)$: $G(r,r,E)=G(z,E)=G(E)\varphi^2(z)$.
The last equality is because all electrons are on the lowest dimensional energy
 in z-direction level and have the same wave function $\varphi^2(z)$.
The magnetic field $H$ is assumed to be strong so that the electron interaction
with impurities $V({\bf r})$ does not mix the wave functions of
electrons on different Landau levels. Hence, the density of states
$\rho (E)$ depends only on the quantity $\omega \equiv E-\varepsilon_0-(n+1/2)\omega_c$,
where
$\varepsilon_0$ is the energy of the dimensional quantization along the $z$-axis and
$\omega_c$ is the cyclotron energy. The interaction potential of
electrons with point-like impurities is
\begin{equation}
\label{2V}
V({\bf r})=u_0\sum_i \delta (x-x_i)\delta (y-y_i)~
\delta (z-z_i)
\end{equation}
Here $x,y,z,x_i,y_i,z_i$ are the electron and impurity coordinates.
The scattering length
\begin{equation}
\label{2}
a=-\frac{m}{2\pi\hbar^2}u_0.
\end{equation}

The Green's function $G(\omega)$ is related to the unperturbed
Green's function
$G_0(\omega)$ of clean electron system by the well-known formula
\cite{AGD}
\begin{equation}
\label{3}
G(\omega )=\frac{1}{G^{-1}_0(\omega )-\Sigma (\omega)},
\qquad\qquad
G_0(\omega )=\frac{1}{\omega }.
\end{equation}
The function $\Sigma (\omega)$ (\ref{3})
is \cite{By,AGD,TA}.
 \begin{equation}
\label{4}
\Sigma (\omega) =u_0n_{imp}
\int_{0}^{z_{\infty}} \frac{\varphi^2_0(z)dz}
{1-\frac{u_0}{2\pi l^2_H}
\varphi^2_0(z)G(\omega)}
\end{equation}
This expression corresponds to the summation of all diagrams
without intersection of impurity lines \cite{TA} and subsequent
averaging over the impurity positions.
The formulas (\ref{3},\ref{4}) are valid only for point-like
impurities (\ref{2V}). It is convenient to determine the reduced wave function
$\varphi_{*}$ of perpendicular-to-layers electron motion:
\begin{equation}
\label{5}
\varphi^2_0(z)=\frac{1}{z_0}\varphi^2_{*}(z_{*}), \qquad\qquad
z_{*}\equiv \frac{z}{z_0}, \qquad\qquad
\int \varphi^2_{*}(z_{*}) ~dz_{*}=1,
\end{equation}
where $z_0$ -- is the characteristic scale of the function $\varphi_0(z)$.
From (\ref{2}), (\ref{4}), (\ref{5}) we get
\begin{equation}
\label{6}
\Sigma (\omega )=-G^{-1}(\omega ) C_0 J(\omega); \qquad\qquad
J(\omega)\equiv \int_0^{z_{\infty }/z_0} \frac{\varphi^2_{*}(z_{*})dz_{*}}
{\varphi^2_{*}(z_{*})-\frac{1}{\omega_0 G(\omega )}}
\end{equation}
where $C_0\equiv 2\pi l^2_Hz_0n_{imp}; \quad
\omega_0\equiv -\frac{a}{z_0}\omega_c$.

\medskip

{\bf 3.} At the beginning we consider the first order in the impurity concentration.
Thus we replace the exact Green's function in formulas
 (\ref{4}), (\ref{6}) by the unperturbed Green's function $G_0=1/\omega $.
 This approximation means the summation of all diagrams
corresponding to the scattering on one impurity.
From (\ref{3}), (\ref{6}) one sees, that the Green's function $G$ formally has
a pole at $\omega =0$:
\begin{equation}
\label{7}
G=\frac{1}{\omega [1+C_0J(\omega )]}.
\end{equation}
However, as follows from (\ref{6}), at $\omega\to 0$ the integral over
$z_{*}$ diverges, that leads to the small value of the residue at this
pole:
\begin{equation}
\label{8}
\frac 1\pi~Im G(\omega) =
\delta (\omega) \frac{1}{1+C_{\infty}}; \qquad\qquad
C_{\infty}=C_0\frac{z_{\infty}}{z_0}
\end{equation}
For $z_{\infty}=\infty $ the residue is zero and 
the infinite degeneracy of the Landau levels
is lifted completely even in the first order in the impurity
concentration. The density of states
$\rho (\omega)$ is determined by the function
$J (\omega)$ (\ref{6})
\begin{equation}
\label{9}
\rho (\omega) = \frac{-C_0ImJ (\omega)}
{\pi\omega [ (1+C_0ReJ(\omega ))^2+
C^2_0Im^2J(\omega)]}
\end{equation}

First, we consider the model case when the reduced wave function $\varphi_{*}$
has the form
\begin{equation}
\label{10}
\varphi^2_{*}(z_{*})=e^{-z_{*}} \quad
\mbox {\rm при} \quad z_{*}>0
\end{equation}
\begin{displaymath}
\varphi^2_{*}(z_{*})=0  \quad
\mbox {\rm при} \quad z_{*}<0 .
\end{displaymath}
This wave function corresponds to the boundary condition at $z=0$ for narrow and deep potential well.
The function $J(\omega )$  (\ref{6}) becomes
\begin{equation}
\label{6J}
J(\omega ) =\ln \frac{\omega-\omega_0}{\omega} .
\end{equation}
Assuming that there is only one type of repulsive impurities, 
from (\ref{9}), (\ref{6J}) we obtain
\begin{equation}
\label{11}
\rho (\omega) =\frac{C_0}
{\omega \biggl [ \biggl ( 1+C_0 \ln \frac{\omega_0-\omega}{\omega}
\biggr )^2+C^2_0\pi^2 \biggr ] } .
\end{equation}
The density of states $\rho (\omega)$ is determined by the cut (not a pole)
of the Green's function $G (\omega)$. Using the definition of $\omega_0$ (\ref{6}),
we get the criterion when the electron-impurity interaction $V({\bf r})$ (\ref{2})
is weak: $a\ll z_0$.
Hence, if the scattering length of electron on impurity is
less than the characteristic scale of the quantization along $z$-axis, the one Landau level
approximation can be used, and the density of states from different Landau levels does not
intersect. From (\ref{11}) it follows that in the wide region of $\omega$
the density of states is inversely proportional to $\omega$:
\begin{equation}
\label{12}
\rho (\omega)=\frac{C_0}{\omega}, \quad
 C_0\ln~\frac{\omega_o}{\omega}\ll 1
\end{equation}

Up to the terms $\sim C_0^2$ the integral
\begin{equation}
\label{14}
\int~\rho (\omega )~d\omega \cong \int\limits^{\omega_0}_0
\frac{C_0d\omega}{\omega  [ 1+C_0~\ln~
\frac{\omega_0-\omega}{\omega}  ]^2}\cong 1 .
\end{equation}
Hence, after switching on the electron-impurity interaction  the
total number of electron states does not change, while the
$\delta$-shape pike of the unperturbed density of states is absent.
This means that at any small value of $\omega$ there is an impurity
with coordinate $z_i=z_0~\ln \frac{\omega_0}{\omega}\gg z_0$,
that lift the Landau level degeneracy. 
These statements do not depend on the exact shape
of the wave function $\varphi (z)$. For more physical $\varphi (z)$ (\ref{1}) the density of
states $\rho (\omega)$ is given by the formula similar to (\ref{11}) at 
$\omega\ll \omega_0$ up to the terms of the order
$\ln\ln\frac{\omega_0}{\omega}$. The dependence $\varphi
(z)\sim e^{-z/z_0}$  at $z\gg z_0$ is important here. Moreover, there is always a wide region
 $\omega$, where $\rho (\omega)$
is inversely proportional to $\omega$. For example, for the wave function 
of harmonic oscillator
 $\varphi_{*}(z_{*})\sim e^{-z^2_{*}}$ from (\ref{9}) and (\ref{6}) 
 at $\omega \ll \omega_0$ we get:
\begin{equation}
\label{15}
\rho (\omega)=\frac{C_0}{2\omega  (
\ln~\frac{\omega_0}{\omega} )^{1/2}
[ 1+C_0
( \ln~\frac{\omega_0}{\omega}
)^{1/2}
]^2} .
\end{equation}
In the case of strong electric field that clamps the electrons to the interface of two media:
$\varphi^2_{*}(z_{*})\sim e^{-z^{3/2}_{*}}$ and from (\ref{9})  at
$\omega\ll \omega_0$ we have
\begin{equation}
\label{16}
\rho (\omega) =\frac{2C_0}
{3\omega (\ln \frac{\omega_0}{\omega})^{1/3}
[1+C_0(\ln \frac{\omega_0}{\omega})^{2/3}]^2}.
\end{equation}

So, for the open electron systems the electron-impurity interaction lifts completely the
degeneracy of the Landau levels. The density of states $\rho (\omega)$ in the first order in the impurity
concentration is given by Eq. (\ref{11})-(\ref{16}). The generalization to the case of several
types of impurities is trivial. For example, the analogue of (\ref{12})
is
\begin{equation}
\label{17}
\rho (\omega)=\frac{C_{-}}{\omega} \quad
\mbox {\rm при} \quad \omega > 0
\end{equation}
\begin{displaymath}
\rho (\omega) = \frac{C_{+}}{|\omega |} \quad
\mbox {\rm при} \quad \omega < 0
\end{displaymath}
where $C_{-}$ and $C_{+}$ are the total concentration of repulsive and attractive impurities.

\medskip

{\bf 4.} The formulas $\rho (\omega)$ (\ref{11})-(\ref{16}) are
valid at
\begin{equation}
\label{lim}
\omega_0 e^{-1/C_0}\ll \omega , \ \omega_0 -\omega \gg e^{-1/C_0}.
\end{equation}
Beyond this region the terms of the next orders in $C_0$ are important, since they
are multiplied by the large logarithm $\ln [(\omega_0 -\omega )/\omega ]$.
To study the Green's function in a wider region over $\omega $ let us consider the self-consistent approximation,
given by formulas (\ref{3}), (\ref{4}), (\ref{6}).
After substitution of (\ref{10}) to (\ref{6}) and (\ref{3}) we get an equation for $G$:
\begin{equation}
\label{Geq}
G=\frac{1}{\omega } [1-C_0 \ln (1-\omega_0 G)] .
\end{equation}
At $\omega \ll \omega_0, \ \omega_0 G\gg 1$ the equation (\ref{Geq})
can be simplified:
\begin{equation}
\label{Geq1}
G=\frac{-C_0}{\omega } [ \ln (-\omega_0 G)-1/C_0]
 =\frac{-C_0}{\omega } \ln \left( -\frac{e \omega_N G}{C_0} \right) ,
\end{equation}
where we have introduced the new scale of energy $\omega_N \equiv \omega_0 \exp
(-1/C_0) C_0/e$.
It is convenient to define the new function $y\equiv -\omega G/ C_0 $. The equation (\ref{Geq1})
takes the form:
\begin{equation}
\label{Geq2}
y=\ln ( y \ \omega_N /\omega )+1,
\end{equation}
From this equation it follows that $y$ depends only on one variable: $y=y(\omega /\omega_N )$.
Let us separate the real and imaginary parts of this function: $y=A+iB$. From (\ref{Geq2})
we get the system of two equations on real functions $A$ and
$B$, that can be simplified to:
$$
A=B/\tan B
$$
\begin{equation}
\label{Geq3}
e^{B/\tan B}\frac{\sin B}{B} = \frac{\omega_N e}{\omega }
\end{equation}
The last equation determines the density of states
$\rho (\omega )=C_0 B/\pi \omega $. We need the solution
$B(\omega /\omega_N)\geq 0$.
The function $B(\omega /\omega_N)$ monotonically increase from zero at
$\omega = \omega_N$ to
$B=\pi $ at $\omega /\omega_N\rightarrow\infty$.
Near $\omega = \omega_N$
the function $B\approx \sqrt{2(\omega -\omega_N)/(\omega_N)}$.
At $\omega_N \ll \omega \ll \omega_0$ up to double logarithmic terms in
$\omega /\omega_N$ the function $B =\pi [1- 1/\ln (\omega /\omega_N)]$.
So, at $\omega \ll \omega_0$ the density of states is given by the universal function
of $\omega /\omega_N$. At $C_0\ll 1$ the asymptotics of this function are:
$$
\omega \rho (\omega ) /C_0 =\sqrt{2(\omega -\omega_N)/(\omega_N)}/\pi ,
\ at
\omega -\omega_N \ll \omega_N;
$$
\begin{equation}
\label{LimB}
\omega \rho (\omega ) /C_0 = 1- 1/\ln (\omega /\omega_N) , \ at \omega \gg\omega_N.
\end{equation}

Let us consider the region $\vert \omega -\omega_0 \vert \ll \omega_0$,
where in first order on $C_0$ we had singularity of
$Im G_0 (\omega )$.
The solution $G(\omega)$ of self-consistent equation (\ref{3}),(\ref{4})
does not have this singularity and monotonically falls to zero at 
$\omega=\omega_x=\omega_0 (1+x_0)$,
where
$x_0$ is the solution of algebraic equation
$x_0=C_0[\ln(1/C_0)+1+\ln(1+x_0)]$. At $C_0\ll 1$ we have
$$x_0\approx C_0[\ln(1/C_0)+1+C_0\ln(1/C_0)].$$
In the vicinity of $\omega = \omega_x$
$B(\omega )\approx \sqrt{2(\omega_x-\omega )[1/(\omega_0 C_0) - 1/\omega_x]}$.
On Fig. 1 we plot the reduced density of states
$ (\omega / C_0 ) \rho (\omega /\omega_0) = B/\pi $ for three different values of impurity concentration.
One sees that at $\omega \ll \omega_0 $ all three plots, while being shifted with $\omega_N$, have
universal shape. The obtained solution give the right behavior of
the density of states $\rho (\omega )$
until it goes to zero, i.e. in the region
 $\omega_N < \omega < \omega_x$. Beyond this interval there are ''tails''
 of the density of states, that are presumably exponentially small
 and can be found only after taking into account the diagrams with
 intersection of the impurity lines.

In conclusion we remark that the proposed model disagrees with the existing
conception. Some exact results of the theory of electron-impurity
interaction in 2D electron systems are obtained in \cite{7,8,9}.
However, in these works the two-dimensional point-like
impurities were considered, and the interaction potential $V(r)$ (\ref{2V}) had
the form
\begin{equation}
\label{18}
V(r)=u_0\delta (x-x_i) \delta (y-y_i).
\end{equation}
Since the $z$-coordinate does not enter here, this potential corresponds to the $\delta$-shaped thread,
but not to a point-like impurity. Our consideration takes
three-dimensional point-like impurities. If one does not integrate over $z$-coordinate
with the impurity distribution function and the wave function of electrons,
one comes to the unphysical limit of $\delta$-shaped threads.

\medskip

The work was supported by grants RFBR and INTAS N01-0791.

\bigskip

Figure caption: 
The reduced density of states $\omega  \rho (\omega ) / C _{0}$
in logarithmic scale for three different values
of the impurity concentration: $C_0= 0.2$ (solid line), $C_0=0.1$ 
(dot line) and $C_0=0.05$ (dash lie).
For each curve the value $\omega_N$ is marked. Near $\omega=\omega_N$ 
the reduced density of states is an universal function of $\omega /\omega_N $. 

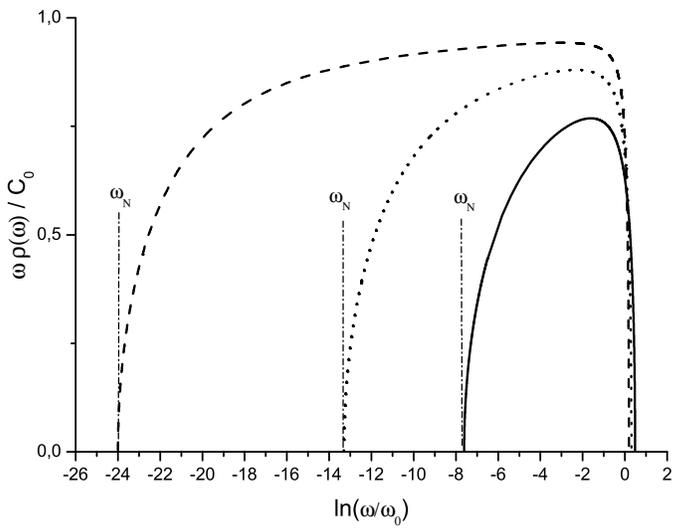
\end{document}